\documentclass[trackchanges,twocolumn]{aastex701}
\usepackage{amsmath}
\usepackage{silence}
\usepackage{hyperref}
\usepackage[noabbrev,capitalize]{cleveref}
\usepackage[mathlines]{lineno}
\usepackage{silence}
\usepackage{booktabs}
\usepackage{microtype}
\usepackage{natbib}
\usepackage{tikz}
\usepackage{graphicx}
\usepackage{makecell}
\usepackage{orcidlink}
\usepackage[caption=false]{subfig}
\usepackage{glossaries-extra}
\glsdisablehyper
\providecommand{\DIFdel}[1]{}

\let\tablenum\relax  \usepackage{siunitx}
\usepackage{placeins}
\usepackage[gobble=auto]{pythontex}
\setpythontexcontext{textwidth=\the\textwidth, linewidth=\the\linewidth, columnwidth=\the\columnwidth}
\begin{pycode}
    from lyla.pythontex_liaison import PythontexLiaison
    from lyla.figure import (
        create_latex_figure_from_bokeh_figure,
        create_latex_figure_from_bokeh_layout,
        create_latex_figure_from_manually_sized_bokeh_layout
    )
    PythontexLiaison.pytex = pytex
\end{pycode}

\usepackage[american]{babel}

\usepackage{etoolbox}
\makeatletter
\patchcmd\H@refstepcounter{\protected@edef}{\protected@xdef}{}{}
\makeatother

\newcommand{\eg}{\textit{e.g.}}
\newcommand{\ie}{\textit{i.e.}}
\newcommand{\approximately}{\raisebox{0.5ex}{\texttildelow}}

\newcommand{\vacuumModelSpeedup}{400}

  \newcommand{\nnColor}{red}
\newcommand{\pmColor}{blue}  \setabbreviationstyle[acronym]{long-short-user}

\newacronym{NN}{NN}{neural network}
\newacronym{LC}{LC}{light curve}
\newacronym{ML}{ML}{machine learning}
\newacronym{MCMC}{MCMC}{Monte Carlo Markov Chain}
\newacronym{MdNSE}{MdNSE}{median normalized squared error}
\newacronym[user1={\eg, \citealp{lecun2015deep}}]{DNN}{DNN}{deep neural network}
\newacronym[user1={\citealp{krizhevsky2012imagenet}}]{CNN}{CNN}{convolutional neural network}
\newacronym[user1={\citealp{panaretos2019statistical}}]{WD}{WD}{Wasserstein distance}
\newacronym[user1={\citealp{massey1951kolmogorov}}]{KS}{KS}{Kolmogorov–Smirnov statistic}
\newacronym[user1={\citealp{menendez1997jensen}}]{JSD}{JSD}{Jensen–Shannon divergence}
\newacronym[user1={\citealp{glorot2011deep}}]{ReLU}{ReLU}{rectified linear unit}
\newacronym{NS}{NS}{neutron star}
\newacronym{EoS}{EoS}{equation of state}
\newacronym{SVF}{SVF}{static vacuum field}
\newacronym{FF}{FF}{force-free} 
\begin{document}
    \title{Pioneering High-Speed Pulsar Parameter Estimation Using Convolutional Neural Networks}

\newcommand{\gsfcAffiliationString}{NASA Goddard Space Flight Center, Greenbelt, MD 20771, USA}
\newcommand{\umdAffiliationString}{Department of Astronomy, University of Maryland, College Park, MD 20742, USA}
\newcommand{\suraAffiliationString}{Southeastern Universities Research Association, Washington, DC 20005 USA}
\newcommand{\cresstAffiliationString}{Center for Research and Exploration in Space Science and Technology, NASA/GSFC, Greenbelt, MD 20771, USA}
 
\author[orcid=0000-0001-8472-2219]{Greg Olmschenk}
\affiliation{\gsfcAffiliationString}
\affiliation{\umdAffiliationString}
\email{greg@olmschenk.com}

\author[orcid=0000-0001-9225-4136]{Emily Broadbent}
\affiliation{\gsfcAffiliationString}
\affiliation{\cresstAffiliationString}
\affiliation{\suraAffiliationString}
\affiliation{Department of Astronomy and Astrophysics, University of California San Diego, La Jolla, CA 92093, USA}
\email{emily.b473@gmail.com}

\author[orcid=0000-0003-1080-5286]{Constantinos Kalapotharakos}
\affiliation{\gsfcAffiliationString}
\email{konstantinos.kalapotharakos@nasa.gov}

\author[orcid=0009-0002-9936-4041]{Wendy F. Wallace}
\affiliation{\gsfcAffiliationString}
\affiliation{\cresstAffiliationString}
\affiliation{\suraAffiliationString}
\email{wendy.f.wallace@nasa.gov}

\author[orcid=0000-0001-6284-2842]{Thibault Lechien}
\affiliation{\gsfcAffiliationString}
\affiliation{\cresstAffiliationString}
\affiliation{\suraAffiliationString}
\affiliation{Max Planck Institute for Astrophysics, Karl-Schwarzschild-Straße 1, 85748 Garching bei München, Germany}
\email{thibaultlechien@gmail.com}

\author[orcid=0000-0002-9249-0515]{Zorawar Wadiasingh}
\affiliation{\gsfcAffiliationString}
\email{zorawar.wadiasingh@nasa.gov}

\author[orcid=0000-0002-7435-7809]{Demosthenes Kazanas}
\affiliation{\gsfcAffiliationString}
\affiliation{\umdAffiliationString}
\email{demos.kazanas-1@nasa.gov}

\author[orcid=0000-0001-6119-859X]{Alice Harding}
\affiliation{Theoretical Division, Los Alamos National Laboratory, Los Alamos, NM 87545, USA}
\email{ahardingx@yahoo.com}

\received{2025-06-20}
\revised{2025-08-22}
\accepted{2025-09-03}
\published{2025-09-26}
\submitjournal{The Astronomical Journal}     \begin{abstract}
    Accurate thermal emission models of neutron stars are essential for constraining the dense matter equation of state. However, incorporating realistic magnetic field structures is computationally prohibitive, severely constraining feasible parameter space exploration. In this work, we develop a \gls{NN} emulator to generate model thermal bolometric X-ray light curves of millisecond pulsars with multipolar magnetic fields. We assess the NN's predictive and computational performance across a broad parameter space. We find that for a \gls{SVF} model, the NN provides a \textgreater{}\vacuumModelSpeedup{} times speedup. We integrate this NN emulator into a \gls{MCMC} framework to replace the computationally expensive physical model during parameter exploration. Applied to PSR J0030+0451, this approach allows the MCMC to reach equilibrium in \approximately{}1 day on 4000 cores, where with the original physical model alone it would have taken more than a year on the same hardware. We compare posterior distributions by running equivalent MCMC iterations with both the NN and the physical model, evaluate differences in distributions when continuing the physical model MCMC from the NN MCMC equilibrium state, and assess variations in posterior distributions resulting from NNs trained on datasets of different sizes. Our NN architecture is agnostic to the underlying physics of the physical model and can be trained for any other physical model, opening many previously intractable avenues of analysis. The NN speed remains the same regardless of the complexity of the physical model it was trained to emulate, allowing greater speedups for more complex physical models.
\end{abstract}
     \section{Introduction}
Understanding \glspl{NS} and their \gls{EoS} has important implications for cold dense matter and fundamental physics. The millisecond pulsar PSR J0030+0451~\citep{somer2000new, d2000bologna} has provided strong evidence for multipolar magnetic fields through modeling of its X-ray phase-folded \gls{LC}~\citep{miller2019psr, riley2019nicer,2019ApJ...887L..23B}. These X-rays arise from deep heating by bombardment from charges at magnetic polar caps; PSR J0030+0451 exhibits a distinctly asymmetric LC demanding departures from a centered dipole field. To constrain the underlying multipolar magnetic field structure of PSR J0030+0451, \cite{kalapotharakos2021multipolar} investigated magnetic field structures that comprised offset dipole-plus-quadrupole components
using \gls{SVF} and force-free global magnetosphere models, uncovering several degenerate parameter solutions for bolometric LCs. However, deriving detailed \gls{MCMC} posterior parameter distributions of these models is prohibitively computationally expensive. In this work, we develop a \gls{NN} surrogate model that replaces the physical models during MCMC parameter exploration, achieving an unprecedented speedup in modeling light curves from pulsar magnetospheres. The resulting acceleration enables the accurate and comprehensive exploration of parameter space that was previously not possible. This also sets the stage for future surrogate models that incorporate additional physics and parameters required to advance beyond bolometric LCs.     \section{Physical model}\label{sec:physical_model}

The methods used in this work for the \gls{SVF} physical model as well as the \gls{MCMC} implementation, are those used in \citet{kalapotharakos2021multipolar}. The method employs a ray-tracing approach utilizing the Kerr metric that integrates photon trajectories from a distant observer image plane to the stellar surface hot spots, with Doppler boosting~\citep[e.g.,][]{2014ApJ...792...87P}. The model assumes that emission originates from hot spots coinciding with the polar caps, \ie, the regions on the stellar surface corresponding to the origin of open\footnote{Open field lines are considered those that cross the light cylinder.} magnetic field lines. We utilize the photon library, simplified atmosphere model, and \gls{LC} normalization from \citet{kalapotharakos2021multipolar}, with the stellar mass, radius, and observer angle fixed to the highest loglikelihood values provided by \citet{riley2019nicer}. Throughout the remainder of this work, unless otherwise specified, the ``physical model'' refers to the \gls{SVF} bolometric thermal X-ray \gls{LC} model.

There are 11 physical parameters input to the physical model (and the \gls{NN}). For the dipole component, the parameters are the 3 Cartesian offset position coordinates relative to the \gls{NS} center $\{x_D, y_D, z_D\}$, the inclination angle $\alpha_D$, and the azimuthal direction of its moment $\phi_D$. For the quadrupole component, the parameters are the 3 Cartesian offset position coordinates $\{x_Q, y_Q, z_Q\}$, the inclination angle $\alpha_Q$, and the azimuthal direction of its moment $\phi_Q$. An additional parameter is the relative strength of the quadrupole moment $B_Q/B_D$ at a distance equal to the \gls{NS} radius.
     \section{Simulated Dataset}\label{sec:simulated_dataset}

The simulated dataset used to train and (partially) evaluate the \gls{NN} consists of parameter sets and the corresponding model X-ray \glspl{LC} produced using the \gls{SVF} model. Our dataset contains \approximately{}$5\times10^8$ \glspl{LC}. It is worth noting that, in the context of an 11-dimensional parameter space, $5\times10^8$ data points still represent a relatively sparse sampling (\ie, \approximately{}$6$ points per dimension for a uniform grid). These \glspl{LC} are generated in two steps.

First, \glspl{LC} are generated for approximately $8\times10^6$ random parameter sets uniformly distributed across the entirety of reasonable parameter space. All parameters are sampled from a uniform distribution. Each of $\{x_D, y_D, z_D, x_Q, y_Q, z_Q\}$ are selected from $\mathcal{U}(-0.5 r_{*},0.5 r_{*})$. However, an additional constraint ensures that the offsets of the dipole and quadrupole moments do not exceed 0.7 of the stellar radius. $\{\phi_D, \phi_Q\}$ are selected from $\mathcal{U}(0, 2\pi)$. $\{\alpha_Q, \alpha_Q\}$ are selected from $\mathcal{U}(0, \pi)$. $B_Q/B_D$ is selected from $\mathcal{U}(0.5, 10.5)$. Each parameter is independently sampled, and each parameter set is independent. The generation of this set of \glspl{LC} took \approximately{}1 day on 4000 cores (\approximately{}40 core seconds per \gls{LC}).

Second, for each parameter set, beyond the original \gls{LC} produced by that set, we generate 63 additional cases by considering a full $360^\circ$ rotation in 64 steps, corresponding to the 64 phase bins of the bolometric NICER X-ray \glspl{LC}. This is achieved by rotating the offsets and magnetic moment orientations and applying a corresponding shift to the \gls{LC}. This approach takes advantage of rotational symmetry to augment the dataset by a factor of 64 without recalculating the physical model \glspl{LC}.

The total size of the simulated dataset is approximately $64 \times 8 \times 10^6 \approx 5\times10^8$ samples. From this dataset, $10^5$ samples are used for validation and another $10^5$ samples for testing. The remainder is used for training. When splitting the dataset, no rotation-augmented versions of a \gls{LC} from the training dataset were included in the validation or test dataset, and vice versa. Except for grouping rotation augmented \glspl{LC} into the same split of the data, splitting between training, validation, and testing is random.
     \section{Neural network pipeline}\label{sec:neural_network_pipeline}

\begin{figure}[t]
    \centering
    \includegraphics[width=\columnwidth]{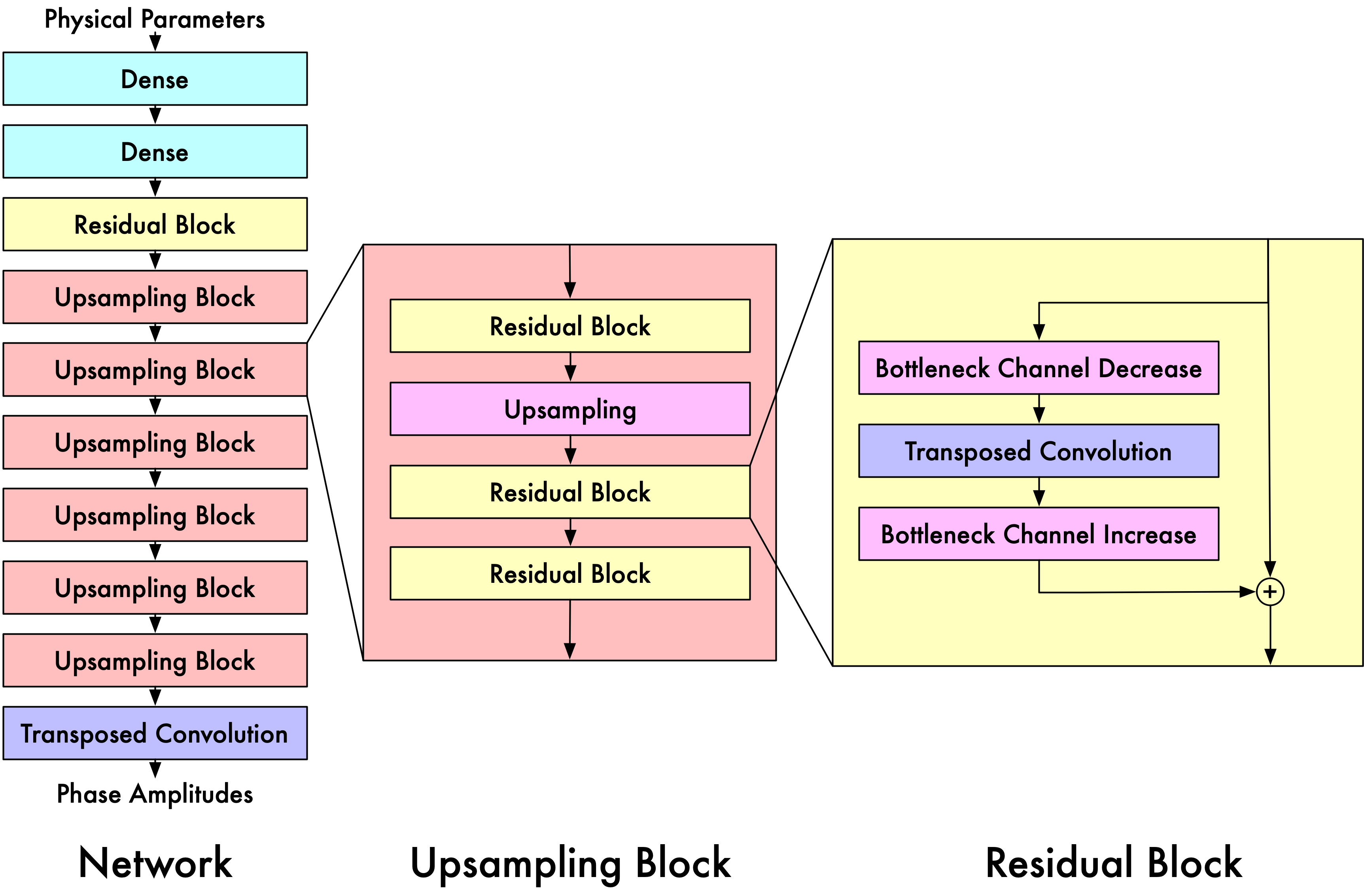}
    \caption{An overview diagram of the neural network model used in this work. The network is a ResNet-like model, except 1D and transposed.}
    \label{fig:cura_network_overview_diagram}
\end{figure}

The network is a ResNet-like model~\citep{he2016deep}, except 1D and transposed. \cref{fig:cura_network_overview_diagram} shows an overview diagram of the \gls{NN} model used in this work. Notably, the architecture includes no structure or mechanisms explicitly designed for \gls{SVF} modeling and can be trained to emulate other physical models. Our full \gls{NN} code, including all network architecture details, is open-source and available at \url{https://github.com/golmschenk/haplo}~\citep{olmschenk_2025_14814679}. The network contains 60 trainable layers with a total of \approximately{}$2\times10^7$ trainable parameters.

For the \gls{NN}, briefly, in addition to the standard advantages of a \gls{CNN} (\eg, weight sharing), a \gls{CNN} is apt for this task as feature locality is important in phase-folded \glspl{LC}, and these networks explicitly take this into account. ResNet architectures are designed to overcome a problem of accuracy degardation of deep networks caused by the difficulty in optimizing deep layers~\citep{he2016deep}. ResNet architectures are robust structures that have been thoroughly tested on an enormous number of tasks. The specific architecture we built for this task was designed based on prior experience working with other \gls{LC} applications. Notably, the architecture includes no structure or mechanisms explicitly designed for \gls{SVF} modeling and can be trained to emulate other physical models.

Though the primary components of the architecture are not uncommon in fields dominated by \glspl{NN}, they are not yet commonplace in the domain of astrophysics, so here we provide a brief, intuitive overview of the architecture’s key elements to help guide readers less familiar with these methods. 

Dense (a.k.a, fully-connected, feed-forward, multilayer-perceptron) layers are layers where each neuron has a connection to each input. Transposed convolutional layers can be seen as doing something near equivalent to the reverse of a standard convolutional layer. The upsamplings simply upsample the data with no learned components. The bottleneck increase and decrease layers are layers designed explicitly to reduce the size of feature space during the transposed convolution, and this reduces the size of the weights tensor of that operation. Finally, ResNet architectures provide ``residual connections'' (the source of ``Res'' in the name), where the output of one block can (in some sense) skip past the layers of the next block. This can be seen as allowing each block to have something close to direct access to the final output. In turn, this allows each block to only need to try to solve the residual error remaining after the previous blocks attempted their solutions to the problem. These connections also provide a ``shortcut'' by which individual blocks can be directly optimized, avoiding the degradation problem noted above. We direct the reader to \citet{he2016deep} for a more complete description.

Here we provide some brief details of the network structure. For full details, please see the code at \url{https://github.com/golmschenk/haplo}. We use an AdamW optimizer~\citep{loshchilov2017decoupled}, all activations are Leaky ReLUs~\citep{he2015delving}, and all trainable weights are initialized using the Kaiming Uniform initialization~\citep{he2015delving}. We use a \gls{MdNSE} metric as the loss metric. Details of this metric and the reasoning for its use are detailed in Section~\ref{sec:evaluation}. The network takes as input the 11 parameters described in Section~\ref{sec:simulated_dataset} and outputs 64 phase amplitudes. For the network layers, the two initial dense layers both have 400 features. The first residual block has 128 features. Each of the upsampling blocks have a consistent number of features for each residual block within it. With that, the number of features for the upsampling blocks are [512, 512, 1024, 1024, 2048, 2048]. The bottlenecks reduce the number of features by a factor of 4, such that each transposed convolutional layer within a residual block has 1/4th the number of features that the residual block has.

To prevent computational numerical issues the inputs and outputs to the network are normalized by the standard deviation and mean values in our training dataset. That is, before being passed to the network, each parameter normalized based on the mean and standard deviation of that parameter in the training dataset. Similarly, the predicted light curve amplitudes are rescaled to the original units. This rescaling is applied before calculating the loss, and so it does not have an impact on how the metric used to evaluate the fit. It is intended only to prevent computational numerical issues due to large numbers internal to the network and their relation to network initializations.

While the trained \gls{NN}'s \gls{LC} generation is orders of magnitude faster than the physical model's, it is important to note that there is an upfront training time cost before the \gls{NN} can be used for this purpose. During our various experiments, the longest total training time a \gls{NN} received was \approximately{}15 days when being trained across 128 Nvidia A100 GPUs on NASA's \textit{Pleiades}. While this training cost is not insignificant, as we will see in the Evaluation Section, the computational cost is much lower than running a single \gls{MCMC} of the physical model to convergence. Furthermore, the model only needs to be trained once, and then can be used for any number of viable solutions (such as the degenerate solutions presented in \citealt{kalapotharakos2021multipolar}). Future models that include stellar mass, radius, and period parameters will additionally be able to be used for any number of \glspl{NS}. From this point of view, the upfront training cost approaches comparative insignificance. Reasonable results can also be obtained with far less training time than that given above. For example, when training with our smaller 5M dataset, we reach training convergence in about 1 day, and, as shown below, the model trained in this way produces a similar posterior distribution for the observed data. Subsequent training times may also be reduced by fine-tuning from existing trained models rather than from random initial conditions.
     \section{Evaluation}\label{sec:evaluation}

We assess the \gls{NN}'s reconstruction performance using two complementary approaches. The first evaluates the \gls{NN}'s ability to emulate the \gls{LC} production of the physical model. The second evaluates how faithfully the posterior of the MCMC using the NN model matches the posterior of the MCMC using the physical model.

\subsection{Light curve reconstruction accuracy on simulated data}
To compare the \gls{NN}'s reconstruction performance, we compare the \gls{NN} predicted \gls{LC} against the \gls{LC} generated by the physical model using the following \glsxtrlong{MdNSE} metric,
\begin{equation}
    \text{MdNSE} = \frac{ \sum_{i} \left( y_{i} - \hat{y_{i}} \right)^2 }{ \text{median}(\hat{y})^2 }
    \text{,}
\end{equation}
where $y$ is the predicted \gls{LC} and $\hat{y}$ is the physical model \gls{LC}. The \gls{MdNSE} was chosen as a stand-in for $\chi^{2}$ for multiple reasons. First, the simulated data does not have associated error values, which are essential for the computation of $\chi^{2}$. Even assuming Poissonian noise for the $\chi^{2}$ calculations is nontrivial. In our simulations, model amplitudes, $\hat{y}_{i}$, depend on an assumed observer-plane distance, unrelated to any specific astrophysical object or observational dataset. Thus, while relative \glspl{LC} amplitude ratios between different parameter sets remain distance-independent, their absolute amplitudes, set by the shared distance scaling, cause $\chi^{2}$ values, computed under Poisson assumptions (where uncertainties scale as the square root of amplitudes), to scale differently. This results in a distance-dependent signal-to-noise ratio, introducing complexities that are difficult to standardize for training purposes. Furthermore, the \glspl{LC} have amplitudes that vary by orders of magnitude. Using $\chi^{2}$ would treat higher-amplitude \glspl{LC} as inherently more important, leading to evaluation biases. In contrast, the MdNSE metric, by normalizing using the median of the physical model \gls{LC}, treats all \glspl{LC} more or less equally, focusing on their shape rather than amplitude. The \gls{NN} is trained to minimize the $\log_{10}(\text{MdNSE})$. The \gls{MdNSE} metric is used only in the context of evaluating the \glspl{LC} produced by the \gls{NN} to those from the simulated dataset during the training of the \gls{NN}. When running the MCMC to estimate the posterior of observed data, using either the physical model or \gls{NN} model, a standard $\chi^{2}$ metric comparison is employed, where $\chi^{2}$ is computed using the observational uncertainties associated with the measured light curve.

\begin{figure}
    \centering
    \begin{pycode}
        from evaluation_resources.random_and_real_grouped_light_curve_comparison import create_random_and_real_grouped_light_curve_comparison_figure
        create_latex_figure_from_bokeh_layout(bokeh_layout=create_random_and_real_grouped_light_curve_comparison_figure(),
                                              latex_figure_path='evaluation_resources/random_and_real_grouped_light_curve_comparison.png',
                                              latex_width=r'\columnwidth',
                                              latex_height=r'2\columnwidth')
    \end{pycode}
    \includegraphics[width=\columnwidth]{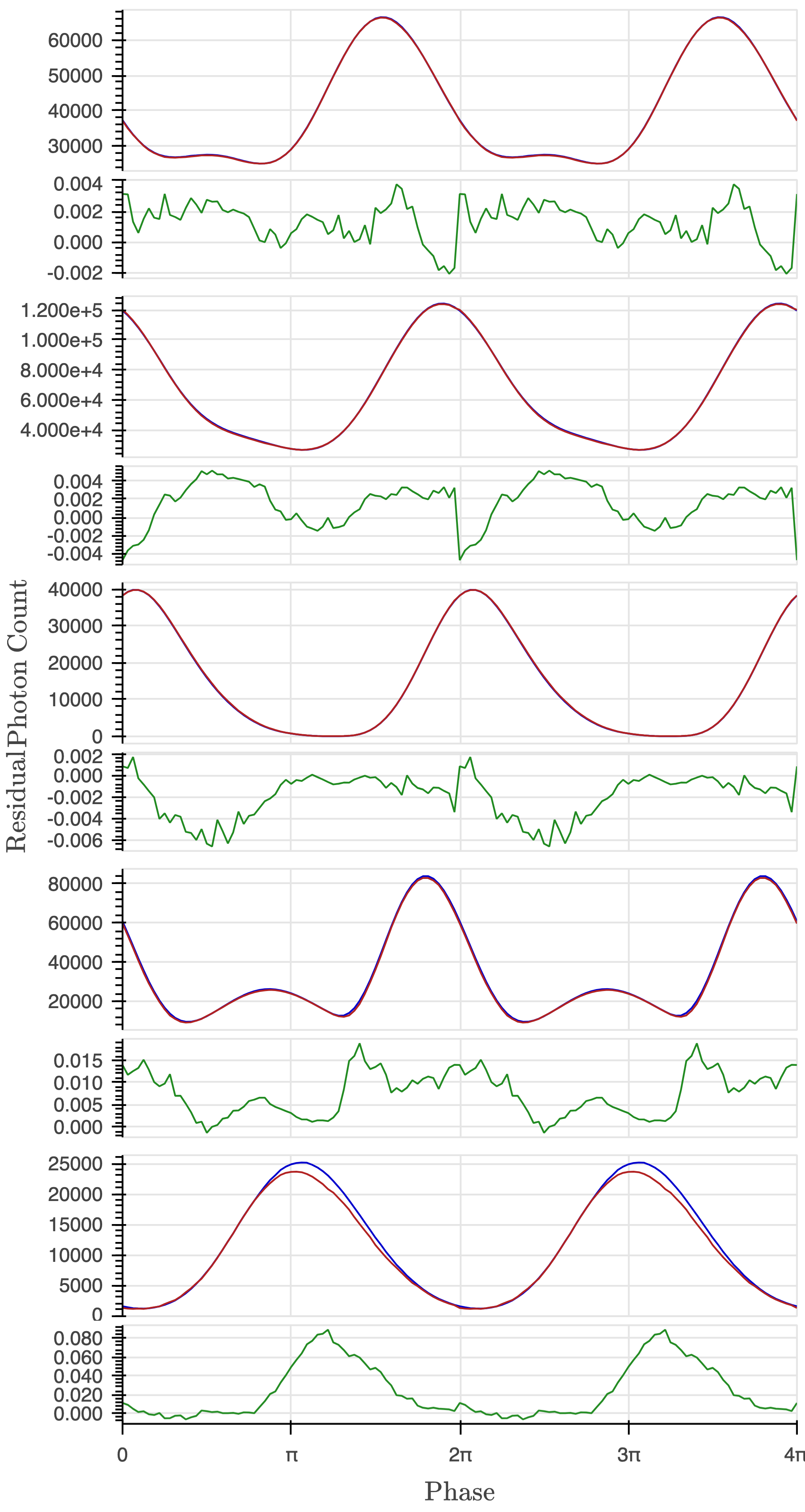}
    \caption{A comparison of light curves produced from the physical model (blue) and the best neural network model (red). From top to bottom, the cases shown are those at quantile $0.0225$, $0.159$, $0.5$, $0.841$, and $0.9775$ of fits sorted by MdNSE. The green line beneath each comparison shows the residual as a fraction of the maximum amplitude of the physical model \gls{LC}.}
    \label{fig:random_and_real_grouped_light_curve_comparison}
\end{figure}

\cref{fig:random_and_real_grouped_light_curve_comparison} presents a representative comparison of \glspl{LC} produced by the physical model and the \gls{NN} model. For an impartial sampling, the \glspl{LC} shown are those at various quantile thresholds of fitting performance with the bottom \gls{LC} corresponding to the 97.75th percentile and the top to the 2.25th percentile of MdNSE. While \cref{fig:random_and_real_grouped_light_curve_comparison} provides an intuitive understanding of the quality of the fits, a more detailed quantitative analysis of the fits is we examine the distribution of MdNSE.

\begin{figure}
    \centering
    \begin{pycode}
        from evaluation_resources.random_and_real_qi_distribution_comparison import create_random_and_real_qi_distribution_comparison_figure
        create_latex_figure_from_bokeh_layout(bokeh_layout=create_random_and_real_qi_distribution_comparison_figure(),
                                              latex_figure_path='evaluation_resources/random_and_real_qi_distribution_comparison.png',
                                              latex_width=r'\columnwidth',
                                              latex_height=r'1.1\columnwidth')
    \end{pycode}
    \includegraphics[width=\columnwidth, height=1.1\columnwidth]{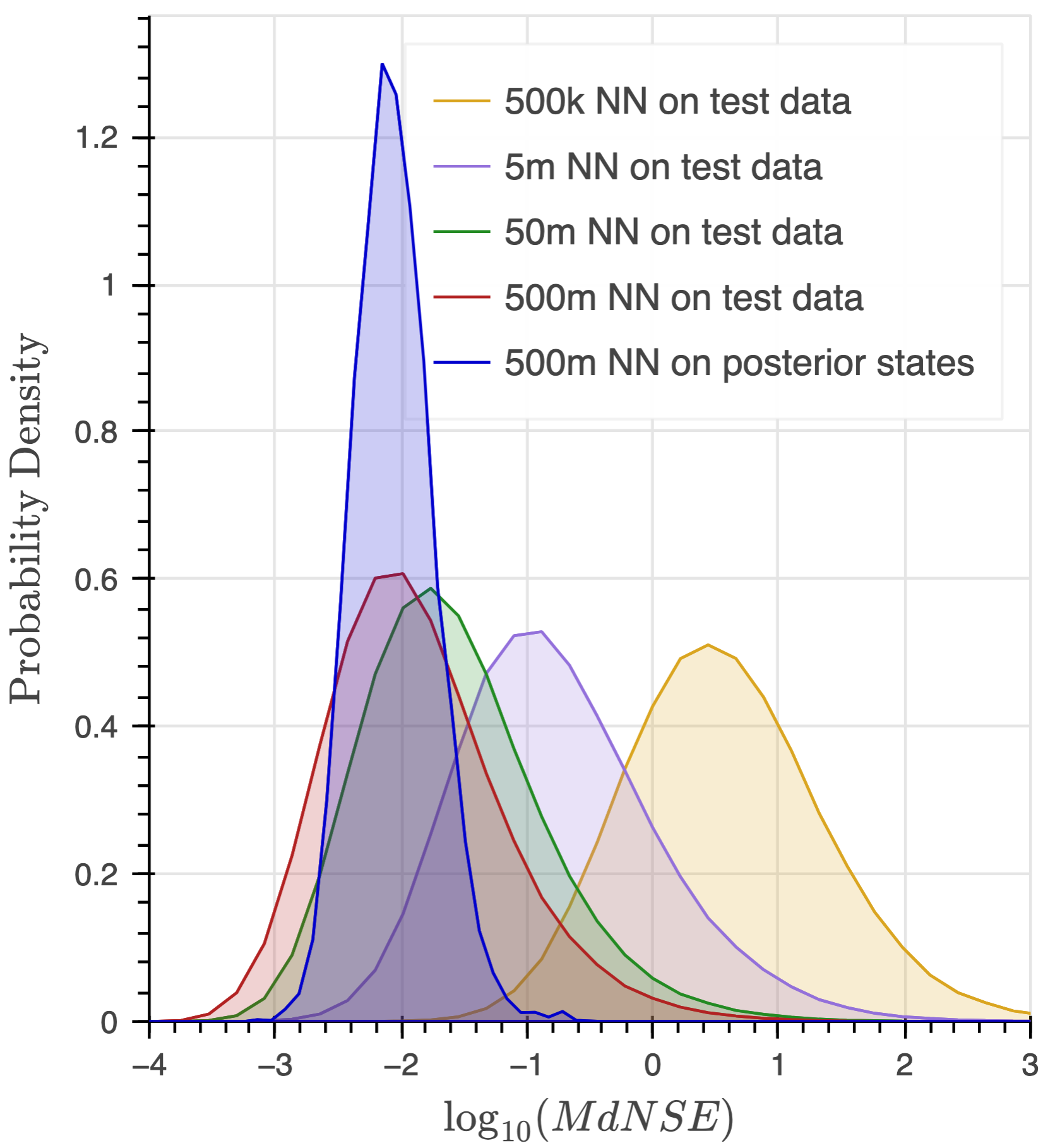}
    \caption{A comparison of the MdNSE distributions of light curve reconstruction on the test dataset and for simulated data using parameter states from an MCMC for PSR J0030+0451. Distributions are shown for the MdNSE values for neural networks trained with 500M (red), 50M (green), 5M (purple), and 500K (yellow) samples when applied to the test dataset, with the network progressively improving with more training data. The blue distribution shows the MdNSE distribution of the best NN (trained with 500M samples) when applied to the parameter states from an MCMC for PSR J0030+0451 run using the physical model.}
    \label{fig:random_and_real_qi_distribution_comparison}
\end{figure}

\cref{fig:random_and_real_qi_distribution_comparison} shows the distributions of MdNSE values in log space when reconstructing \glspl{LC} from the random test dataset using NNs trained on varying dataset sizes, ranging from 500k to 500M samples. All distributions exhibit an approximate Gaussian distribution shape in $\log(\text{MdNSE})$, with both the mean and median values shifting to lower MdNSE values and the standard deviation decreasing as the training dataset size increases. This trend indicates improved reconstruction accuracy with larger training datasets. The best performing \gls{NN} model is the one trained with 500M samples.

\subsection{Distribution reconstruction accuracy on real data}
\begin{figure*}
    \centering
    \begin{pycode}
        from evaluation_resources.posterior_comparison import create_posterior_comparison_with_best_nn_fitted_model_figure

        create_latex_figure_from_manually_sized_bokeh_layout(bokeh_layout=create_posterior_comparison_with_best_nn_fitted_model_figure(),
                                                             latex_figure_path='evaluation_resources/posterior_comparison_with_best_nn_fitted_model.png')

        from evaluation_resources.metrics import generate_metrics_latex_table

        generate_metrics_latex_table('500M unconverged vs physical model unconverged', 'evaluation_resources/metrics_comparison_with_best_nn_fitted_model.tex')
    \end{pycode}
    \begin{tikzpicture}
        \node[anchor=south west, inner sep=0] (image) at (0, 0) {\includegraphics[width=\textwidth]{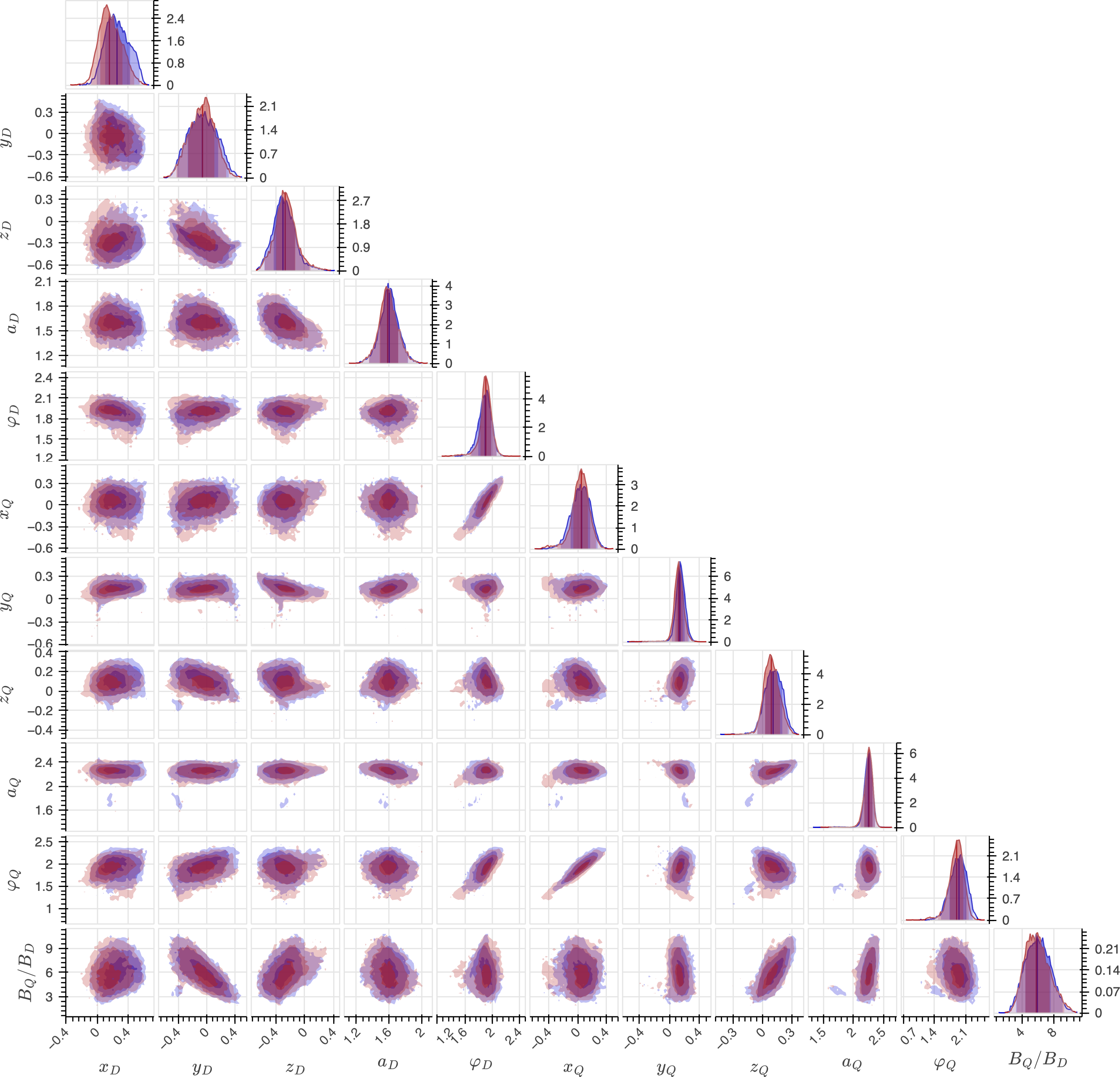}};
        \node[anchor=north east] at (image.north east) {
            \begin{tabular}{lrrrr}
\toprule
Parameter & WD & KS & JSD & CI for median \\
\midrule
$x_{D}$ & 0.584 & 0.223 & 0.219 & 44.06\% \\
$y_{D}$ & 0.096 & 0.059 & 0.108 & 0.81\% \\
$z_{D}$ & 0.161 & 0.080 & 0.096 & 14.70\% \\
$a_{D}$ & 0.093 & 0.055 & 0.083 & 10.22\% \\
$\varphi_{D}$ & 0.178 & 0.108 & 0.135 & 13.63\% \\
$x_{Q}$ & 0.123 & 0.051 & 0.122 & 0.48\% \\
$y_{Q}$ & 0.330 & 0.153 & 0.147 & 30.34\% \\
$z_{Q}$ & 0.217 & 0.117 & 0.123 & 22.68\% \\
$a_{Q}$ & 0.043 & 0.021 & 0.083 & 1.22\% \\
$\varphi_{Q}$ & 0.218 & 0.124 & 0.154 & 18.62\% \\
$B_{Q}/B_{D}$ & 0.071 & 0.041 & 0.088 & 6.52\% \\
\bottomrule
\end{tabular}
         };
    \end{tikzpicture}
    \caption{A comparison of the PSR J0030+0451 posterior distributions from the MCMC with light curves generated by the physical model (blue) and the best neural network model (red). Both MCMCs executed approximately 8192 parallel chains for 8000 steps. The physical model version took \approximately{}5 days on 4000 cores. The neural network version took \approximately{}15 minutes on the same hardware. The contours of the distribution show the credible intervals equivalent to 1-, 2-, and 3-$\sigma$s. Metrics comparing the 1D marginal distributions between the posteriors is shown at the top right. Shown are the Wasserstein distance (WD), the Jensen–Shannon divergence (JSD), and the Kolmogorov–Smirnov statistic (KS). Also shown is the maximum credible interval required to encompass the median of the other distribution (CI for median).}
    \label{fig:posterior_comparison_with_best_nn_fitted_model_figure}
\end{figure*}
\cref{fig:posterior_comparison_with_best_nn_fitted_model_figure} shows a comparison of the posterior distributions for PSR J0030+0451 as estimated via MCMC runs using the physical model to produce \glspl{LC} (\pmColor{}) and the NN model (\nnColor{}). Both MCMC runs started from the same initial conditions as the MCMC run that produced the RV1 solution presented in \citet{kalapotharakos2021multipolar} and were executed with approximately 8192 parallel chains, all iterated for 8000 steps, more than twice the number of steps used in \citet{kalapotharakos2021multipolar}. The physical model MCMC run required 5 days on 4000 Intel Broadwell cores, making further evolution computationally prohibitive. By comparison, on the same hardware, the NN model MCMC run completed the same number of iterations in a small fraction of the time, i.e., \approximately{}15 minutes.

In the table shown within the figure, for the 1D marginalized distributions, the credible interval required to encompass the median value of the distribution from the other method is shown (CI for median). As there are two distributions that contain CIs for the other distribution's median to fall within, the larger of these two values is the one that is given here. From this, we see the median of one distribution is always well within the 68.27\% credible interval (1$\sigma$) of the other, and is below 15\% in most cases. This provides an initial intuitive indication of the NN’s reliability in replicating the posterior results of the physical model.

To more thoroughly quantify the similarity between the distributions, we compute the \gls{WD}, a metric that evaluates the difference between two probability distributions, capturing details beyond the median and standard deviation. The parameter values are first normalized using their modified $Z$-scores, ensuring standardization across the distributions. The mean \gls{WD} is then calculated for the normalized 1D marginal distributions, providing a robust measure of the discrepancy between the NN and physical model results. We also include the \gls{JSD} and \gls{KS} values. However, they show similar trends to WD. For all three metrics, a value of zero would show perfect agreement. However, there is no clear reference value for any of these metrics of what constitutes ``good enough''. That value is both problem and distribution dependent. We hope the ``CI for median'' provides a human-interpretable metric of the agreement of the distributions and the more detailed metrics can be used for specific purposes. Related to the lack of clear reference values, we note that we exclude the p-value for the \gls{KS} metric. As we expect the two distributions are not exactly identical, and we can sample as many points as we wish from the distribution, we are able to arbitrarily scale the p-value.

The distribution comparison metrics for each parameter are displayed within the corner plot, offering a direct visual comparison between the posterior distributions and their quantified differences. The results demonstrate that the NN model closely agrees with the physical model, even for higher-dimensional parameter spaces. Importantly, the spread of the NN-based posterior distributions is consistent with the physical model, suggesting that the NN faithfully reproduces not only the central tendencies but also the variability of the underlying parameter space.

In \cref{fig:random_and_real_qi_distribution_comparison}, we also show the $\log(\text{MdNSE})$ distribution of the \glspl{LC} corresponding to the posterior distribution shown in \cref{fig:posterior_comparison_with_best_nn_fitted_model_figure}. Although this $\log(\text{MdNSE})$ distribution is narrower than the corresponding distribution for the 500M \gls{NN} on all of reasonable parameter space, the two distributions are comparable around the same median value. This suggests that the NN's performance in reproducing the posterior distribution of PSR J0030+0451 is not an outlier and similar performance can be expected for other regions of parameter space.

With the computational speed achieved by the NN implementation, we were able to extend the MCMC explorations to approximately $1.5\times 10^6$ iterations for the same 8192 parallel chains. This longer run converged to a posterior distribution that differed from the one derived from the shorter MCMC runs limited by the speed of the physical model. Importantly, this longer run MCMC resulted in a posterior distribution that stabilized and remained consistent over the last 1 million iterations, providing strong evidence of convergence. Notably, this result, and a subsequent analysis of the physical model MCMC evolution, revealed that the solutions identified in \citet{kalapotharakos2021multipolar} were not yet converged.

\begin{figure*}
    \centering
    \begin{pycode}
        from evaluation_resources.posterior_comparison import create_posterior_comparison_with_converged_pm_continued_from_converged_best_nn_fitted_model_figure
        create_latex_figure_from_manually_sized_bokeh_layout(bokeh_layout=create_posterior_comparison_with_converged_pm_continued_from_converged_best_nn_fitted_model_figure(),
                                                             latex_figure_path='evaluation_resources/posterior_comparison_with_converged_pm_continued_from_converged_best_nn.png')

        from evaluation_resources.metrics import generate_metrics_latex_table

        generate_metrics_latex_table('500M converged vs physical model continued from 500M converged', 'evaluation_resources/metrics_comparison_with_converged_pm_continued_from_converged_best_nn_fitted_model.tex')
    \end{pycode}
    \begin{tikzpicture}
        \node[anchor=south west, inner sep=0] (image) at (0, 0) {\includegraphics[width=\textwidth]{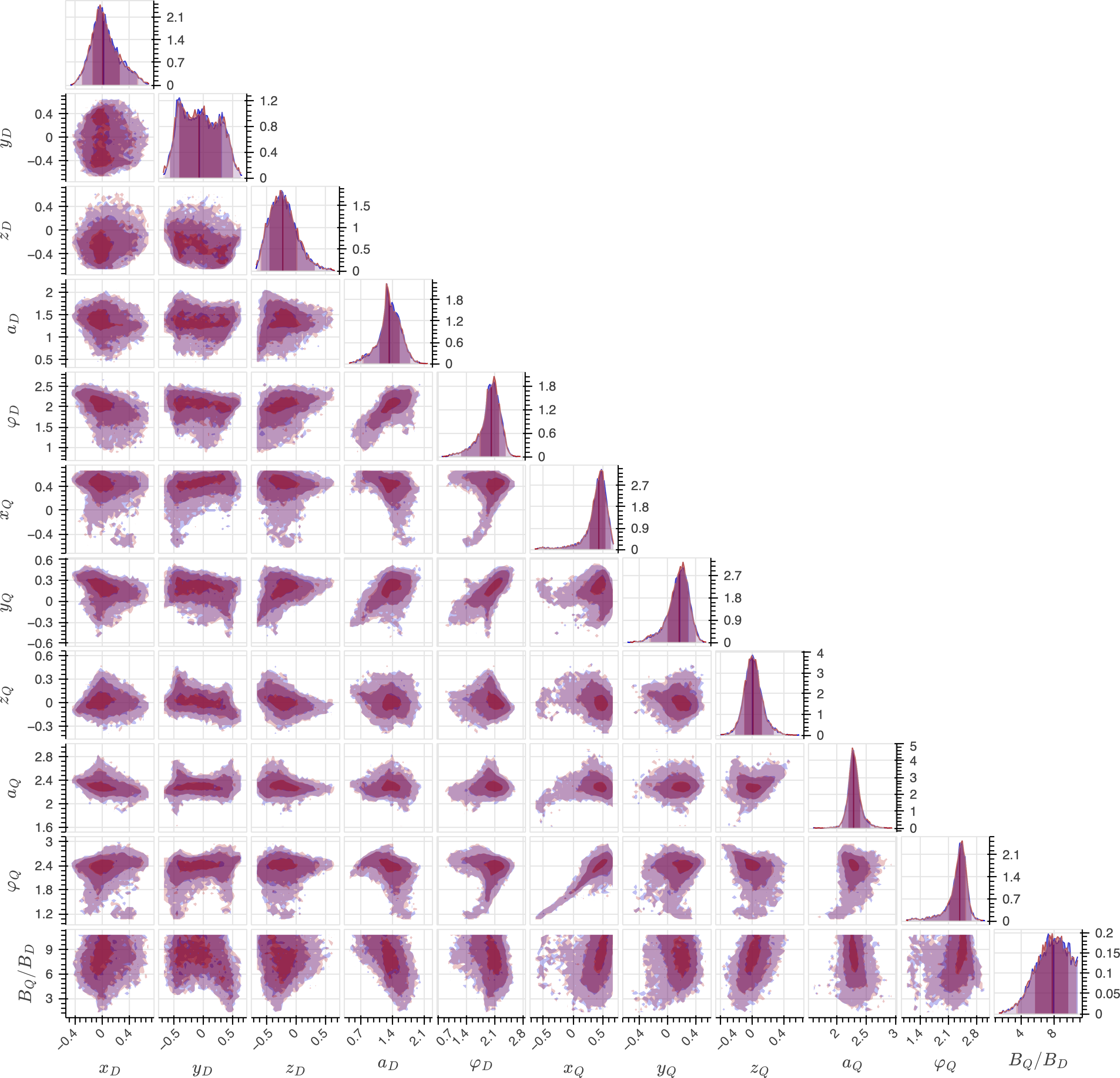}};
        \node[anchor=north east] at (image.north east) {
            \begin{tabular}{lrrrr}
\toprule
Parameter & WD & KS & JSD & CI for median \\
\midrule
$x_{D}$ & 0.020 & 0.009 & 0.095 & 1.15\% \\
$y_{D}$ & 0.015 & 0.007 & 0.098 & 0.62\% \\
$z_{D}$ & 0.019 & 0.009 & 0.099 & 0.36\% \\
$a_{D}$ & 0.029 & 0.012 & 0.098 & 2.20\% \\
$\varphi_{D}$ & 0.028 & 0.008 & 0.085 & 0.53\% \\
$x_{Q}$ & 0.015 & 0.014 & 0.099 & 0.75\% \\
$y_{Q}$ & 0.046 & 0.015 & 0.086 & 0.66\% \\
$z_{Q}$ & 0.019 & 0.010 & 0.084 & 0.80\% \\
$a_{Q}$ & 0.030 & 0.013 & 0.090 & 1.74\% \\
$\varphi_{Q}$ & 0.011 & 0.007 & 0.104 & 0.74\% \\
$B_{Q}/B_{D}$ & 0.020 & 0.013 & 0.097 & 1.72\% \\
\bottomrule
\end{tabular}
         };
    \end{tikzpicture}

    \caption{A comparison of the PSR J0030+0451 posterior distributions from the MCMC with light curves generated by the physical model (blue) and the best neural network model (red). In this case, the MCMC was run to convergence using the neural network model. Then, the physical model MCMC was continued from that converged state to investigate if it would diverge from the converged state found by the neural network MCMC. Using the neural network, the MCMC took approximately 1 day on 4000 cores to reach convergence. To reach the same number of MCMC iterations (and presumably the same convergence) using the physical model would take more than a year on the same hardware. The contours of the distribution show the credible intervals equivalent to 1-, 2-, and 3-$\sigma$s.}
    \label{fig:posterior_comparison_with_converged_pm_continued_from_converged_best_nn_fitted_model_figure}
\end{figure*}

\begin{figure*}
    \centering
    \begin{pycode}
        from evaluation_resources.posterior_comparison import create_posterior_comparison_of_500m_50m_and_5m_figure
        create_latex_figure_from_manually_sized_bokeh_layout(bokeh_layout=create_posterior_comparison_of_500m_50m_and_5m_figure(),
                                                             latex_figure_path='evaluation_resources/posterior_comparison_of_500m_50m_and_5m_figure.png')
    \end{pycode}
    \begin{tikzpicture}
        \node[anchor=south west, inner sep=0] (image) at (0, 0) {\includegraphics[width=\textwidth]{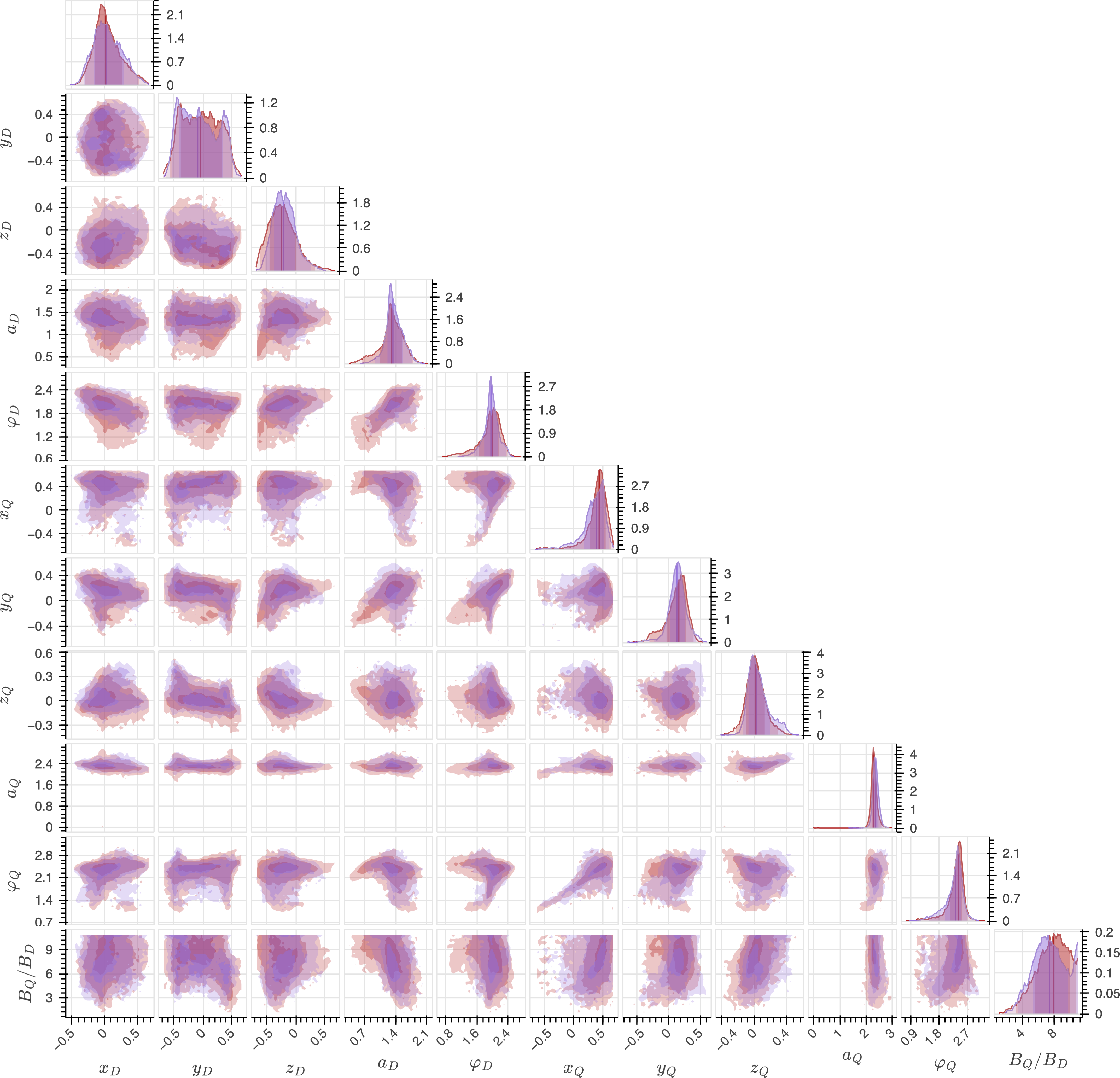}};
        \node[anchor=north east] at (image.north east) {
            \begin{tabular}{lrrrr}
\toprule
Parameter & WD & KS & JSD & CI for median \\
\midrule
$x_{D}$ & 0.122 & 0.042 & 0.081 & 5.74\% \\
$y_{D}$ & 0.156 & 0.052 & 0.091 & 10.30\% \\
$z_{D}$ & 0.323 & 0.110 & 0.152 & 14.84\% \\
$a_{D}$ & 0.579 & 0.122 & 0.163 & 13.51\% \\
$\varphi_{D}$ & 0.784 & 0.095 & 0.184 & 10.18\% \\
$x_{Q}$ & 0.448 & 0.164 & 0.159 & 30.76\% \\
$y_{Q}$ & 0.481 & 0.077 & 0.157 & 8.55\% \\
$z_{Q}$ & 0.473 & 0.092 & 0.152 & 17.40\% \\
$a_{Q}$ & 0.799 & 0.253 & 0.219 & 48.33\% \\
$\varphi_{Q}$ & 0.373 & 0.108 & 0.133 & 20.98\% \\
$B_{Q}/B_{D}$ & 0.235 & 0.103 & 0.115 & 20.70\% \\
\bottomrule
\end{tabular}
         };
    \end{tikzpicture}

    \caption{A comparison of the PSR J0030+0451 posterior distributions from the MCMC run to convergence with light curves generated by neural networks with differing amounts of training data. The networks were trained with 500M (red) and 5M (purple) training samples. The contours of the distribution show the credible intervals equivalent to 1-, 2-, and 3-$\sigma$s.}
    \label{fig:posterior_comparison_of_500m_50m_and_5m_figure}
\end{figure*}

However, due to the computational expense of the physical model, evolving it for an equivalent number of iterations is infeasible, preventing direct comparison of these posterior distributions. To address this limitation, first, we performed a continuation study. Starting from the equilibrium posterior distribution of the best-fit NN, we evolved the MCMC for an additional 4000 iterations using the physical model. The resulting posterior showed no significant variation from the converged NN-derived posterior, providing strong evidence that the NN-derived posterior accurately represents the distribution that the physical model would converge to with further exploration. A comparison of these two posterior distributions is shown in \cref{fig:posterior_comparison_with_converged_pm_continued_from_converged_best_nn_fitted_model_figure}. As a second verification, the posterior distribution generated by the NN trained with 5M samples was compared to those from the NN trained on the full 500M dataset. This comparison is shown in \cref{fig:posterior_comparison_of_500m_50m_and_5m_figure}. The agreement between these distributions indicates that the NNs achieve sufficient accuracy to enable reliable posterior estimation, even with reduced training data. The \gls{NN} trained with 5M samples results in a different level of \gls{LC} reconstruction accuracy (as shown in \cref{fig:random_and_real_qi_distribution_comparison}) and exhibits only weak correlation with the \gls{NN} trained with 500M samples. Yet, the fact that it still produces approximately the same converged posterior distribution as the 500M \gls{NN} model provides additional evidence that NN-based MCMCs effectively reveal the true underlying posteriors. The \gls{NN} trained with 50M samples shows comparable agreement.

\subsection{A new solution from convergence}

As noted above, the MCMC run using the physical model for 8000 iterations does not achieve convergence, whereas the NN-based run extended to \approximately{}$1.5\times 10^6$ iterations does converge. This converged posterior distribution is notably different from the unconverged physical model distribution. While this can be seen by comparing the distributions in \cref{fig:posterior_comparison_with_best_nn_fitted_model_figure} and \cref{fig:posterior_comparison_with_converged_pm_continued_from_converged_best_nn_fitted_model_figure}, we also provide a quantitative comparison of the 1D marginal posteriors in \cref{fig:posterior_1d_comparison_table}, where the 1D marginal median values and 68.3\% credible intervals are listed side-by-side. Notably, the median values for $x_{D}$, $x_{Q}$, $z_{Q}$, $\varphi_{Q}$, and $B_{Q}/B_{D}$ for the \gls{NN} converged distribution lie outside the 1$\sigma$-equivalent credible intervals of the unconverged  physical model distribution. These discrepancies underscore the importance of allowing the MCMC to run to equilibrium in order to accurately capture the true posterior structure and underlying physical meaning.

\begin{table*}
    \centering
    \hspace{-8em}
    \begin{tabular}{lrrrrrrrrrrr}
\toprule
 & $x_{D}$ & $y_{D}$ & $z_{D}$ & $a_{D}$ & $\varphi_{D}$ & $x_{Q}$ \\
\midrule
$\begin{tabular}{@{}l@{}}Unconverged\\physical model\end{tabular}$ & $0.258^{{+}0.180}_{{-}0.145}$ & $-0.057^{{+}0.214}_{{-}0.206}$ & $-0.290^{{+}0.147}_{{-}0.137}$ & $1.603^{{+}0.110}_{{-}0.108}$ & $1.900^{{+}0.084}_{{-}0.097}$ & $0.045^{{+}0.127}_{{-}0.139}$ \\
$\begin{tabular}{@{}l@{}}Converged neural\\network model\end{tabular}$ & $0.018^{{+}0.244}_{{-}0.163}$ & $-0.063^{{+}0.381}_{{-}0.339}$ & $-0.225^{{+}0.251}_{{-}0.213}$ & $1.338^{{+}0.239}_{{-}0.229}$ & $2.019^{{+}0.198}_{{-}0.281}$ & $0.440^{{+}0.111}_{{-}0.168}$ \\
\toprule
& $y_{Q}$ & $z_{Q}$ & $a_{Q}$ & $\varphi_{Q}$ & $B_{Q}/B_{D}$ \\
\midrule
$\begin{tabular}{@{}l@{}}Unconverged\\physical model\end{tabular}$ & $0.147^{{+}0.058}_{{-}0.055}$ & $0.110^{{+}0.088}_{{-}0.089}$ & $2.257^{{+}0.060}_{{-}0.067}$ & $1.943^{{+}0.178}_{{-}0.201}$ & $5.952^{{+}1.611}_{{-}1.555}$ \\
$\begin{tabular}{@{}l@{}}Converged neural\\network model\end{tabular}$ & $0.170^{{+}0.120}_{{-}0.166}$ & $0.008^{{+}0.113}_{{-}0.104}$ & $2.291^{{+}0.103}_{{-}0.085}$ & $2.378^{{+}0.148}_{{-}0.252}$ & $7.834^{{+}1.979}_{{-}2.214}$ \\
\bottomrule
\end{tabular}
     \caption{A comparison of the 1D marginal distributions of the unconverged physical model and the converged neural network model. Shown are the 1D marignal median values and the upper and lower bounds of the 68.3\% credible interval.}
    \label{fig:posterior_1d_comparison_table}
\end{table*}

\begin{figure}
    \centering
    \includegraphics[width=\columnwidth]{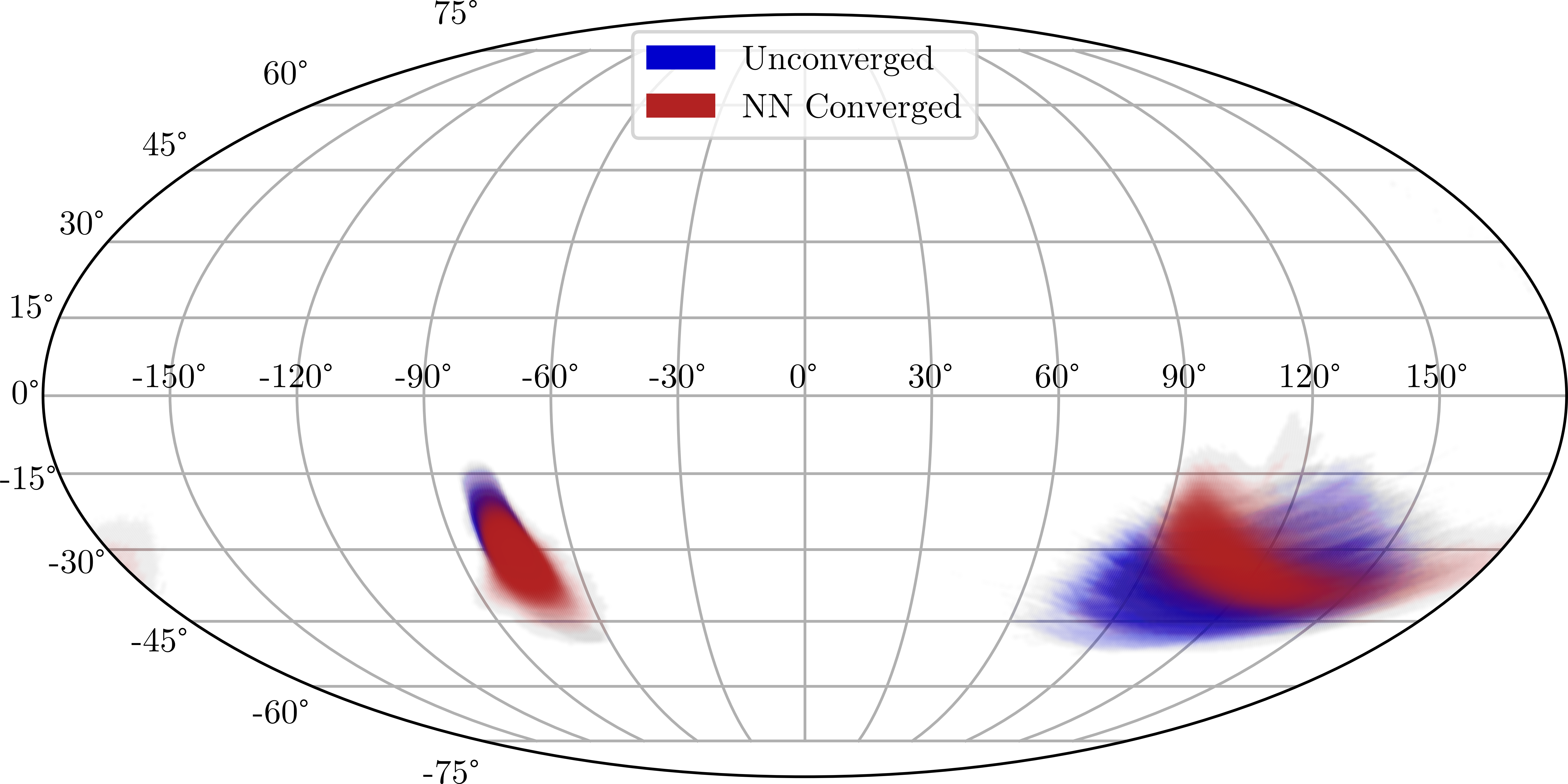}
    \caption{A comparison of the likelihood-weighted hotspot distributions corresponding to the posterior samples before and after the converged solution determined by the NN. The ``Unconverged'' distribution is derived from the posterior of the physical model after 8000 iterations, shown in \cref{fig:posterior_comparison_with_best_nn_fitted_model_figure}. The ``NN Converged'' distribution corresponds to the posterior obtained from the converged neural network model shown in \cref{fig:posterior_comparison_with_converged_pm_continued_from_converged_best_nn_fitted_model_figure}.}
    \label{fig:hotspot_comparison}
\end{figure}

Additionally, in \cref{fig:hotspot_comparison}, we show the likelihood-weighted hotspot distributions projected onto the stellar surface for both the unconverged and converged posteriors. The shifts in hotspot shape and orientation between these two cases illustrate how premature termination of the MCMC can result in a physically distinct interpretation of the system’s emission geometry. This highlights the value of computationally efficient methods like the NN emulator, which make deep convergence accessible in practice.

Finally, it's worth noting that the physical model posterior distribution obtained by continuing the MCMC from the converged \gls{NN} state for an additional \approximately{}4000 iterations does not show any significant deviation from the \gls{NN}-derived distribution.  In both the 1D marginal comparisons and the hotspot maps, the continued physical model posterior distribution yields results nearly identical to those shown in \cref{fig:posterior_1d_comparison_table} and \cref{fig:hotspot_comparison} (see also \cref{fig:posterior_comparison_with_converged_pm_continued_from_converged_best_nn_fitted_model_figure}). This agreement further validates that the NN-derived posterior closely approximates the true equilibrium solution that would be obtained with the physical model, if such a solution were computationally feasible to reach.

     \section{Discussion and Conclusion}\label{sec:conclusion}

This work establishes NNs as a transformative tool for NS parameter inference, overcoming previous computational bottlenecks and enabling extensive parameter space exploration that was previously infeasible. By replacing computationally expensive physical models with a NN emulator, we achieve a speedup of over 400 times for SVF models, reducing MCMC run-times from days to minutes and from years to days. As the SVF models on bolometric LCs themselves are already simplified, this framework bodes favorably for surrogates of more advanced physical models that consider the full parameter space, including mass and radius. The NN's speed enables previously unattainable parameter space exploration and evaluation of posterior distribution convergence, overcoming prior computational limits.

One of the most significant outcomes of this work is the ability to achieve convergence in MCMC explorations, which was infeasible with the SVF physical model in \citet{kalapotharakos2021multipolar}. Extending the MCMC iterations to approximately 1.5 million with the NN, we find that the posterior distributions differ from those obtained in previously reported shorter MCMC runs. Importantly, the posterior stabilized and remained consistent for the final 1 million iterations, providing strong evidence of convergence. Direct comparisons with the physical model for such extended runs remain intractable due to their computational expense. However, validation experiments, including the physical model MCMC continuation from the NN-derived equilibrium state and NNs trained on considerably smaller datasets, indicate that the NN-derived posterior is an accurate representation of the underlying distribution.

The implications of these advancements are even more drastic for force-free (FF) magnetohydrodynamic field models of the global magnetosphere, where the physical model computations are at least three orders of magnitude slower than the SVF case for the same field parameters. Inference using the NN for FF \glspl{LC} incurs no additional computational overhead, as the NN speed depends solely on its architecture. Our architecture has no structures or mechanisms explicitly geared toward any particular physical model, and the same network can be trained to emulate other modeling approaches. Preliminary results for the FF emulation show speedups exceeding $10^6$ times, bringing previously inaccessible analyses to practical time scales. These approaches are under active investigation and will be presented in \citet{lechien2025multiwavelength}, where we expand the parameter space to include stellar mass, radius, and observer angle, further enhancing the scope of this framework.

Beyond computational speed, our extended exploration of the parameter space using the NN-accelerated approach unified certain solutions, such as RV1 and RV2 in \citet{kalapotharakos2021multipolar}, which were initially thought to be distinct. However, other solutions, such as RV3 and RV4, remained distinct, suggesting the existence of multiple local log-likelihood maxima that are relatively prominent, suggesting significant competing regions within the parameter space. These results emphasize the importance of extensive MCMC sampling to explore the entire parameter space thoroughly, a current significant limitation in NICER analyses of millisecond pulsars. While all corresponding reduced $\chi^2$ values remain well below 1, incorporating additional observational constraints, such as gamma-ray \glspl{LC}, is crucial to refining the analysis and identifying dominant solutions, as previously demonstrated in \citet{kalapotharakos2021multipolar}. A comprehensive study along this direction is also going to be presented in \citet{lechien2025multiwavelength}.

In conclusion, this study establishes the NN emulator as a powerful and scalable surrogate for computationally expensive physical models in \gls{NS} parameter inference. The demonstrated speedup not only enhances the feasibility of MCMC-based analyses for SVF models but also paves the way for applications such as FF or other physical models, where computational gains are even more immense. By enabling extensive parameter space explorations and achieving robust convergence, this approach provides a more comprehensive understanding of \gls{NS} properties. These innovations significantly advance efforts to constrain the dense matter equation of state, opening new avenues of research in the field and positioning NNs as an indispensable tool for the future of \gls{NS} analysis.
     \section{Acknowledgments}

The material is based upon work supported by NASA under award numbers 80GSFC21M0002, 80GSFC24M0006, 80NSSC21K1999, 22-ADAP22-0142, 22-TCAN22-0027, 21-ATP21-0116, and 22-FERMI22-0035.     \bibliography{bibliography}
\bibliographystyle{aasjournal}
 \end{document}